\documentclass[a4paper,12pt]{article}
\usepackage{amssymb,amsfonts,amsmath,mathtext,cite,enumerate,float}
\usepackage{amssymb,amsthm,latexsym,euscript,amscd, authblk}
\usepackage{braket}

\usepackage[english]{babel}
\usepackage[cp1251]{inputenc}

\title{Non-commutative graphs in the Fock space over one-particle Hilbert space}

\author[1,2,3]{G.G. Amosov\thanks {gramos@mi-ras.ru}}

\author[4]{A.S. Mokeev\thanks {alexandrmokeev@yandex.ru}}

\affil[1] {Steklov Mathematical Institute of Russian Academy of Sciences,
	ul. Gubkina 8, Moscow 119991, Russia}
\affil[2] {Moscow Institute of Physics and Technology, 9 Institutski per., Dolgoprudny, 141701, Russia}
\affil[3]{Institute of Mathematics, Ufa Federal Research Center of Russian Academy of Sciences, 112 Chernyshevskii st., Ufa, 450077, Russia}
\affil[4] {Chebyshev Laboratory of St. Petersburg State University, 14th Line V.O. 29B, St. Petersburg 199178, Russia}

\date{}

\begin{document}

\maketitle

\begin{abstract}{ In the present paper we continue our study of non-commutative operator graphs in infinite-dimensional spaces. We consider examples of the non-commutative operator graphs generated by resolutions of identity corresponding to the Heisenberg-Weyl group of operators acting on the Fock space over one-particle state space. The problem of quantum error correction for such graphs is discussed.   }
\end{abstract}

keywords: non-commutative operator graphs, covariant resolution of identity, symmetric Fock space, quantum anticliques.

\section{Introduction}

Let $H$ and $B(H)$ be a Hilbert space and the space of all bounded linear operators on $H$ respectively. A non-commutative operator graph is a subspace $\mathcal{V}\subset B(H)$ containing the identity operator and closed under operator conjugation. These objects appear in the theory of quantum error correction. In \cite{Knill1997, Knill2000,Bennet} was established the theory of error-correcting codes for a quantum noise. The existence of a code for the channel determines the ability of transmitting  quantum information perfectly. Each quantum channel set the unique non-commutative operator graph $\mathcal{V}$ such that the subspace $S\subset H$ is a quantum error-correcting code for the channel iff the orthogonal projection $P_S$ on $S$ satisfies the equality $P_S\mathcal{V}P_S=\mathbb{C}P_S$. The projection $P_S$ is called a quantum anticlique. In this setting the non-commutative operator graph is a quantum analogue of the confusability graph for a classical communication channel \cite {Duan}.

Since \cite{amo} a series of papers devoted to the study of operator graphs linearly generated by covariant resolutions of identity was published. For such graphs the sufficient condition to have a quantum anticlique was obtained\cite{amosovmokeev1}.
In the finite-dimensional case some examples for various groups were presented\cite{amosovmokeev1,amosovmokeev2,amosovmokeev3}. As a development of these results we constructed the non-commutative operator graph in infinite-dimensional Hilbert space corresponding to the dynamics of the coupled quantum oscillator. The maximal anticlique was found for this model\cite{AMP}. 

The theory of coherent states and squeezed coherent states that emerged through pioneering work \cite {gla, sud} 
is a very well known instrument in quantum optics \cite {Klauder, Manko}. We use the theory of coherent states to construct examples of quantum anticliques for the operator graph corresponding to the coupled quantum oscillator. The famous Glauber-Surdashan equality \cite{gla,sud} is extremely useful for our needs. It should be noted that coherent states have some generalizations\cite{Perelom}. One of such generalizations could be given through exponential vectors in the symmetric Fock space used in quantum field theory to describe the states of a system with variable number of bosonic particles.

In section 2 we give a description of symmetric Fock space over one-particle Hilbert space. In section 3 we derive examples of non-commutative operator graphs in the Fock space over Hilbert space of arbitrary finite dimension. For such graphs we construct anticliques. The similar techniques is used in \cite{AMP}.

\section{Exponential vectors in the symmetric Fock space}

Let $a^{\dag}$ and $a$ be the standard bosonic creation and annihilation operators.
Given $\alpha \in {\mathbb C}$ a coherent state $\ket {\alpha }$ is a solution to the equation
$$
a\ket {\alpha }=\alpha \ket {\alpha }.
$$
Introducing the displacement operator 
$$
D(\alpha )=\exp(\alpha a^{\dag}-\alpha ^*a)
$$
we obtain
$$
D(\alpha )\ket {0}=\ket {\alpha }
$$
and
$$
D(\alpha )D(\beta )=e^{(\alpha \beta ^*-\alpha ^*\beta )/2}D(\alpha +\beta ).
$$
The famous Glauber-Sudarshan equality 
\begin{equation}\label{GS}
\frac {1}{\pi }\int \limits _{\mathbb C}\ket {\alpha }\bra {\alpha }d^2\alpha =I
\end{equation}
can be rewritten in the form
\begin{equation}\label{iden}
\frac{1}{\pi}\int \limits D(\alpha )Q_{\beta }D(\alpha )^*d^2\alpha =I,
\end{equation}
where $Q_{\beta }=\ket {\beta }\bra {\beta }$ is an arbitrary coherent state $\beta \in {\mathbb C}$.

This techniques can be extended to any dimension. Consider the symmetric Fock space over one-particle 
separable Hilbert space $H$
$$
F(H)=\{{\mathbb C}\Omega \}\oplus H\oplus \dots \oplus H^{\otimes _s^n}\oplus \dots , 
$$
where $H^{\oplus _s^n}$ is a symmetrized tensor product of $n$ copies of $H$. Given $f\in H$ an exponential vector $e(f)\in F(H)$
is determined by the formula
$$
e(f)=\sum \limits _{n=0}^{+\infty }\frac {f^{\otimes n}}{\sqrt {n!}},
$$
the linear envelope of exponential vectors is dense in $F(H)$. Let us define the annihilation and creation operators $a(f),a^{\dag }(g)$ as follows:
$$
a(f)e(g)=(f,g)e(g),\ a^{\dag }(f)e(g)=\frac {d}{dt}e(g+tf)|_{t=0}.
$$ 
Then, $a(f)$ and $a(g)$ satisfy the canonical commutation relations
$$
a(f)a^{\dag }(g)-a^{\dag }(g)a(f)=(f,g)I,
$$
$$
a(f)a(g)-a(g)a(f)=0,
$$
$$
a^{\dag}(f)a^{\dag }(g)-a^{\dag }(g)a^{\dag }(f)=0.
$$
Given $f\in H$ define a unitary Heisenberg--Weyl operator by the formula
\begin{equation}\label{HW}
W(f)=\exp(a^{\dag }(f)-a(f)).
\end{equation}
The set of Heisenberg--Weyl operators satisfies the condition
$$
W(f)W(g)=e^{-iIm(g,f)}W(f+g)
$$
Thus, (\ref {HW}) determines a projective unitary representation of the Abelian group $(H,+)$. 
Since
$$
W(f)a(g)W(f)^{\dag}=a(g)+(g,f)I
$$
we get
$$
W(f)e(0) =e^{-\frac {||f||^2}{2}}e(f).
$$
Hence
$$
W(f)e(g)=e^{\frac {||g||^2}{2}}W(f)W(g)e(0)=e^{\frac {||g||^2}{2}}e^{-iIm(g,f)}W(f+g)e(0)
$$
$$
=e^{-\frac {||f||^2}{2}-iIm(g,f)}e(f+g).
$$
(Note that $e(0)=\Omega $.)
Suppose that we have a subset $K\subset H$ considered as a measurable space with some measure $\mu$. Our goal is to obtain the conditions on the orthogonal projection $Q$ and $\mu $ under which
\begin{equation}\label{main}
	\int \limits _{K}W(f)QW(f)^*d\mu (f)=I.
\end{equation}
Equation (\ref {main}) ensures us that the following subspace of $B(F(H))$
$$
\mathcal{V}=\overline{span}\{W(f)QW(f)^*,f \in K\}
$$
is an operator graph.

Consider a composite system $H=\mathfrak {H}\oplus \mathfrak {K}$. In the case,
$F(H)$ and $F(\mathfrak {H})\otimes F(\mathfrak {K})$ are unitary equivalent. The corresponding map is given by the formula
\begin{equation}\label{isom}
U(e(f\oplus g))=e(f)\otimes e(g),\ f,g\in \mathfrak {H},\ g\in\mathfrak{K}.
\end{equation}
If $H=\{{\mathbb C}f\}$ is one-dimensional and $||f||=1$, then given an $\alpha \in {\mathbb C}$ the exponential vector $e(\alpha f)\in F(H)\equiv F({\mathbb C})$ can be identified with
a unnormalized coherent state $e^{\frac {|\alpha |^2}{2}}\ket {\alpha }$. In this case, the Weyl operator $W(\alpha f)$
should be identified with the displacement operator $D(\alpha )$.

\section{Non-commutative operator graphs in the symmetric Fock space}

Let $H$ be the $n$-dimensional Hilbert space with the fixed orthonormal basis $f_k,\ 1\le k \le n$. Then, a linear operator
$$
U_n: F(H) \rightarrow F(\mathbb{C})^{\otimes^{n}}
$$
defined by the formula
$$
U_n(e(\alpha _1 f_1\oplus\dots\oplus \alpha _n f_n))=e^{\frac {1}{2}{\sum \limits _{j=1}^n|\alpha _j|^2}}\ket {\alpha _1}\otimes \dots \otimes \ket {\alpha _n}
$$
fulfils a unitary equivalence between $F(H)$ and $F(\mathbb{C})^{\otimes^{n}}$. 
Consider an arbitrary unitary $n\times n$ matrix 
$$
\Phi=\{\Phi_{jk}, \ 1\le j,k \le n \}.
$$
We also need the inverse unitary matrix
$$
\Phi ^{-1}=\{\Phi _{kj}^*\}.
$$
Let us define a positive linear operator $Q_\Phi :F({\mathbb C})^{\otimes ^n}\to F({\mathbb C})^{\otimes ^n}$ by the formula
$$
Q_{\Phi}=\frac {1}{\pi }\int \limits _0^{2\pi }d\theta \int \limits _0^{+\infty }r d r  \ \bigotimes_{j=1}^{n}\ket {r e^{i\theta }\Phi_{j,1}}\bra {r e^{i\theta }\Phi_{j,1}}.
$$

{\bf Proposition 1.} {\it $Q_{\Phi}$ is the orthogonal projection.}

Proof.

Consider a unitary operator ${\mathcal U}:F(H)\to F(H)$
with the matrix $\Phi ^{-1}$ in the basis $(f_k)$. Then,
$$
U_n{\mathcal U}U_n^*\left (\bigotimes_{j=1}^{n}\ket {r e^{i\theta }\Phi_{j,1}}\right )=\ket {re^{i\theta}}\otimes \ket {0}^{\otimes ^{n-1}}.
$$
Hence,
$$
U_n{\mathcal U}U_n^*Q_{\Phi }U_n{\mathcal U}^*U_n^*=I_{F({\mathbb C})}\otimes \ket {0}\bra {0}^{\otimes ^{n-1}}
$$
in virtue of (\ref {GS}). 
$\Box $

Let us take a set of indices
$$
R=\{r_j\ge 0,\ 1\le j \le n \},
$$
$$
\Theta=\{2\pi \ge \theta_j\ge 0,\ 1\le j \le n \}.
$$
and define a unitary operator $D_{R,\Theta ,\Phi }: F({\mathbb C})^{\otimes ^n}\to F({\mathbb C})^{\otimes ^n}$
by the formula
$$
D_{R,\Theta,\Phi}=\bigotimes_{j=1}^{n} D\left (\sum\limits_{k=2}^{n}e^{i\theta_{k}}r_{k}\Phi_{j,k}\right ).
$$

{\bf Theorem 1.} {\it 
	$$
	\frac {1}{\pi^{n-1} }\int \limits _{0}^{2\pi}\int \limits _{0}^{+\infty}\dots
	\int \limits _{0}^{2\pi}\int \limits _{0}^{+\infty}
    D_{R,\Theta,\Phi}Q_{\Phi}
	D_{R,\Theta,\Phi} \prod_{k=2}^{n}r_kdr_kd\theta_k =I.
	$$
}

{\bf Proof.}

Denote
$$
g_k=\bigoplus _{j=1}^n\Phi _{j,k}f_j,\ 1\le k\le n.
$$
Then, we have
$$
\frac {1}{\pi^{n-1} }\int \limits _{0}^{2\pi}\int \limits _{0}^{+\infty}\dots
\int \limits _{0}^{2\pi}\int \limits _{0}^{+\infty}
U_{n}^{*}D_{R,\Theta,\Phi}Q_{\Phi }
D_{R,\Theta,\Phi}U_{n} \prod_{k=2}^{n}r_kdr_kd\theta_k 
$$
$$
=\frac {1}{\pi^{n} }\int \limits _{0}^{2\pi}\int \limits _{0}^{+\infty}\dots
\int \limits _{0}^{2\pi}\int \limits _{0}^{+\infty}
W(r_{n} e^{i\theta{n}}g_n)\dots W(r_2e^{i\theta_{2}}g_2)W(r_1e^{i\theta _1}g_1) \ket{\Omega}\bra{\Omega}
$$
$$
W(r_1e^{i\theta _1}g_1)^{*}W(r_2e^{i\theta_{2}}g_2)^{*} \dots W(r_{n} e^{i\theta{n}}g_n)^{*}
\prod_{k=1}^{n}r_k  dr_kd\theta_k\equiv \tilde Q.
$$
Since $(g_k)$ are pairwise orthogonal we can define a unitary operator ${U}_{\Phi }:F(H)\to F({\mathbb C}^{\otimes ^n})$ by the formula
\begin {equation}\label{U} 
{U}_{\Phi }(e(\beta _1g_1\oplus \dots \oplus \beta _ng_n))=e^{\frac {1}{2}\sum \limits _{k=1}^n|\beta _k|^2}\ket {\beta _1}\otimes \dots \otimes \ket {\beta _n}.
\end{equation}
It follows that
$$
{U}_{\Phi }\tilde Q\tilde {U}_{\Phi }^*=\frac {1}{\pi ^n}\int \limits _{{\mathbb C}^n}\left (\bigotimes _{k=1}^nD(\beta _k)\right )\ket {0}\bra {0}^{\otimes ^n}\left (\bigotimes _{k=1}^nD^*(\beta _k)\right )\prod _{k=1}^nd^2\beta _k=I 
$$
by means of (\ref{iden}).

$\Box $

{\bf Corollary.} {\it The closure
	$$
	\mathcal {V}_{\Phi }=\overline {span}\{ D_{R,\Theta,\Phi}Q_{\Phi }
	D_{R,\Theta,\Phi}^*,\ \Theta \in [0,2\pi ]^{n-1},\ R\in [0,+\infty)^{n-1}   \}
	$$
	is a non-commutative operator graph.
}

{\bf Theorem 2.} {\it Given $\Phi ,\Gamma \in [0,2\pi ]^{n-1}$ and $X \in (0,+\infty)^{n-1}$
	the projection $P_{\Phi ,\Gamma ,X }$ determined by
	$$
	P_{\Phi ,\Gamma ,X }=U_{\Phi }U_n^*\left(I_{F(\mathbb{C})}\otimes \ket {X_1 e^{i\Gamma_1 }}\bra{X_1 e^{i\Gamma_1 }}\otimes \dots 
	\otimes\ket {X_{n-1} e^{i\Gamma_{n-1} }}\bra{X_{n-1} e^{i\Gamma_{n-1} }}\right)U_nU_{\Phi }^*
	$$
	is a quantum anticlique for ${\mathcal V}_{\Phi }$.}

{\bf Proof.}

 We get
$$
U_nU_{\Phi }^*P_{\Phi ,\Gamma ,X }D_{R,\Theta,\Phi}Q_{\Phi }
D_{R,\Theta,\Phi}P_{\Phi ,\Gamma ,X }U_{\Phi }U_n^*
$$
$$
=\prod \limits _{k=1}^{n-1}|\braket {r_{k+1}e^{i\theta _{k+1}}, X_ke^{i\Gamma _k}}|^2
$$
$$
I_{F({\mathbb C})}\otimes \ket {X_1 e^{i\Gamma_1 }}\bra{X_1 e^{i\Gamma_1 }}\otimes \dots 
\otimes
\ket {X_{n-1} e^{i\Gamma_{n-1} }}\bra{X_{n-1} e^{i\Gamma_{n-1} }}
$$
$$
=C(\Phi ,\Gamma ,X,R,\Theta)P_{\Phi ,\Gamma ,X }
$$
Since $U_\Phi$ and $U_n$ are unitary and $F(\mathbb{C})$ is an infinite-dimensional space so the image of $P_{\Phi ,\Gamma ,X }$ is an infinite-dimensional subspace of $F(H)$. Thus $P_{\Phi ,\Gamma ,X }$ is an anticlique. $\Box $

\section{Conclusion} We constructed non-commutative operator graphs consisting of operators acting on the symmetric Fock space over one-particle state spaces. For such graphs problem of quantum error correction is studied. Quantum anticliques (quantum error correcting codes) are obtained.

\section{Acknowledgments}

Research of A.S. Mokeev is supported by \textquotedblleft Native towns\textquotedblright, a social investment program of PJSC \textquotedblleft Gazprom Neft\textquotedblright.


\begin{thebibliography}{99}


\bibitem{Knill1997} 
E. Knill, R. Laflamme, \textquotedblleft Theory of error-correction codes, \textquotedblright Physical Review A \textbf {55},  900--911 (1997).

\bibitem{Knill2000} 
E. Knill, R. Laflamme, L. Viola, \textquotedblleft Theory of quantum error correction for general noise, \textquotedblright Phys. Rev. Lett. \textbf {84}, 2525 (2000).

\bibitem{Bennet}
C. H. Bennett, D. P. DiVincenzo, J. A. Smolin, W. K. Wootters, \textquotedblleft Mixed state entanglement and quantum error correction,  \textquotedblright Physical Review A \textbf {54}, 3824--3851 (1996).


\bibitem{Duan} 
R. Duan, S. Severini, A. Winter, \textquotedblleft Zero-error communication via quantum channels, noncommutative
graphs and a quantum Lovasz theta function, \textquotedblright IEEE Trans. Inf. Theory. \textbf {59}, 1164--1174 (2013).

\bibitem{amo} 
G.G. Amosov, \textquotedblleft On general properties of non-commutative operator graphs, Lobachevskii Journal of Mathematics, \textquotedblright \textbf {39} (3), 304--308 (2018).

\bibitem{amosovmokeev1} 
G.G. Amosov, A.S. Mokeev, \textquotedblleft On non-commutative operator graphs generated by covariant resolutions of identity, \textquotedblright Quantum Information Processing \textbf {17}, 325 (2018).

\bibitem{amosovmokeev2} 
G.G. Amosov, A.S. Mokeev, \textquotedblleft On non-commutative operator graphs generated by reducible unitary representation of the Heisenberg-Weyl group, \textquotedblright International Journal of Theoretical Physics, doi:10.1007/s10773-018-3963-4; arXiv:1812.02515.

\bibitem{amosovmokeev3}
G.G. Amosov, A.S. Mokeev, \textquotedblleft On linear structure of non-commutative operator graphs, \textquotedblright Lobachevskii J. Math. \textbf{40} (10), 1440--1443 (2019).

 
\bibitem{AMP}
G.G. Amosov, A.S. Mokeev, A.N. Pechen, \textquotedblleft Non-commutative graphs and quantum error correction for a two-mode quantum oscillator,\textquotedblright arXiv:1910.08935.

\bibitem{gla} 
R.G. Glauber, \textquotedblleft Coherent and incoherent states of the radiation field, \textquotedblright Physical Review \textbf {131} (6), 2766--2788 (1963).

\bibitem{sud} 
E.C.G. Sudarshan, \textquotedblleft Equivalence of semiclassical and quantum mechanical descriptions of statistical light beams, \textquotedblright Physical Review Letters \textbf{10} (7), 277--279 (1963).


\bibitem{Klauder} 
J.R. Klauder, E.C.G. Sudarshan, \emph {Fundamentals of quantum optics}, (W.A. Benjamin, INC, 1968).

\bibitem{Manko} 
\emph {The theory of non-classical states of light (ed. V.V. Dodonov, V.I. Man'ko)}, (Taylor \& Francies, 2003).


\bibitem{Perelom}
A. Perelomov, \emph {Generalized Coherent States and Their Applications}, (Springer-Verlag, 1986).


\end{thebibliography}
\end{document}